\documentclass[longauth]{aa} 
\usepackage{graphicx}
\usepackage{natbib}
\usepackage{txfonts}
\begin{document}
   \title{Results of WEBT, VLBA and RXTE monitoring of \object{3C 279}  
   during 2006--2007\thanks{ The radio-to-optical data 
   presented in this paper are stored in the WEBT archive; for questions regarding their availability,
   please contact the WEBT president Massimo Villata ({\tt villata@oato.inaf.it}).}}


   \author{V.~M.~Larionov             \inst{1,2} 
   \and   S.~G.~Jorstad                  \inst{1,21} 
   \and   A.~P.~Marscher                  \inst{21}   
   \and   C.~M.~Raiteri               \inst{ 3}   
   \and   M.~Villata                  \inst{ 3}   
   \and   I.~Agudo                    \inst{ 4}   
   \and   M.~F.~Aller                 \inst{ 5}   
   \and   A.~A.~Arkharov              \inst{ 2}   
   \and   I.~M.~Asfandiyarov          \inst{20}   
   \and   U.~Bach                     \inst{ 6}   
   \and   R.~Bachev                   \inst{ 7}   
   \and   A.~Berdyugin                \inst{28}
   \and   M.~B\"ottcher               \inst{ 8}   
   \and   C.~S.~Buemi                 \inst{16}
   \and   P.~Calcidese                \inst{ 9}
   \and   D.~Carosati                 \inst{10}   
   \and   P.~Charlot                  \inst{11}   
   \and   W.-P.~Chen                  \inst{12}   
   \and   A.~Di Paola                 \inst{13}
   \and   M.~Dolci                    \inst{14}
   \and   S.~Dogru                    \inst{15}   
   \and   V.~T.~Doroshenko            \inst{40,34,41}
   \and   Yu.~S.~Efimov               \inst{34}
   \and   A.~Erdem                    \inst{15}   
   \and   A.~Frasca                   \inst{16}   
   \and   L.~Fuhrmann                 \inst{ 6}
   \and   P.~Giommi                   \inst{36}
   \and   L.~Glowienka                \inst{17}
   \and   A.~C.~Gupta                 \inst{44,18}
   \and   M.~A.~Gurwell               \inst{19}   
   \and   V.~A.~Hagen-Thorn           \inst{1}   
   \and   W.-S.~Hsiao                 \inst{12}   
   \and   M.~A.~Ibrahimov             \inst{20}   
   \and   B.~Jordan                   \inst{40}
   \and   M.~Kamada                   \inst{22}   
   \and   T.~S.~Konstantinova         \inst{ 1}
   \and   E.~N.~Kopatskaya            \inst{ 1}
   \and   Y.~Y.~Kovalev               \inst{6,23}
   \and   Y.~A.~Kovalev               \inst{23}
   \and   O.~M.~Kurtanidze            \inst{24}   
   \and   A.~L\"ahteenm\"aki          \inst{25}   
   \and   L.~Lanteri                  \inst{ 3}   
   \and   L.~V.~Larionova             \inst{ 1}
   \and   P.~Leto                     \inst{37}
   \and   P.~Le Campion               \inst{11}   
   \and   C.-U.~Lee                   \inst{26}
   \and   E.~Lindfors                 \inst{28}
   \and   E.~Marilli                  \inst{16}   
   \and   I.~McHardy                  \inst{27}   
   \and   M.~G.~Mingaliev             \inst{42}
   \and   S.~V.~Nazarov               \inst{34}
   \and   E.~Nieppola                 \inst{25}
   \and   K.~Nilsson                  \inst{28}   
   \and   J.~Ohlert                   \inst{29}
   \and   M.~Pasanen                  \inst{28}
   \and   D.~Porter                   \inst{30}
   \and   T.~Pursimo                  \inst{31}
   \and   J.~A.~Ros                   \inst{32}
   \and   K.~Sadakane                 \inst{22}
   \and   A.~C.~Sadun                 \inst{33}
   \and   S.~G.~Sergeev               \inst{34,41}
   \and   N.~Smith                    \inst{39}
   \and   A.~Strigachev               \inst{ 7}
   \and   N.~Sumitomo                 \inst{22}   
   \and   L.~O.~Takalo                \inst{28}
   \and   K.~Tanaka                   \inst{22}   
   \and   C.~Trigilio                 \inst{16}
   \and   G.~Umana                    \inst{16}
   \and   H.~Ungerechts               \inst{43}
   \and   A.~Volvach                  \inst{35}   
   \and   W.~Yuan                     \inst{18}
}

   \offprints{V.~M.~Larionov, vlar@astro.spbu.ru}

   \institute{
          Astron.\ Inst., St.-Petersburg State Univ., Russia
   \and   Pulkovo Observatory, St.-Petersburg, Russia
   \and   INAF, Osservatorio Astronomico di Torino, Italy                              
   \and   Instituto de Astrof\'{i}sica de Andaluc\'{i}a, CSIC, Granada, Spain
   \and   Department of Astronomy, University of Michigan, MI, USA
   \and   Max-Planck-Institut f\"ur Radioastronomie, Bonn, Germany
   \and   Inst.\ of Astron., Bulgarian Acad.\ of Sciences, Sofia, Bulgaria
   \and   Department of Physics and Astronomy, Ohio Univ., OH, USA
   \and   Oss.\ Astronomico della Regione Autonoma Valle d'Aosta, Italy
   \and   Armenzano Astronomical Observatory, Italy
   \and   Lab. d'Astrophys., Univ.\ Bordeaux 1, CNRS, Floirac, France
   \and   Institute of Astronomy, National Central University, Taiwan
   \and   INAF, Osservatorio Astronomico di Roma, Italy
   \and   INAF, Osservatorio Astronomico di Collurania Teramo, Italy
   \and   COMU Observatory, Turkey
   \and   INAF, Osservatorio Astrofisico di Catania, Italy
   \and   Department of Phys.\ and Astron.\, Univ.\ of Aarhus, Denmark 
   \and   YNAO, Chinese Academy of Sciences, Kunming, China
   \and   Harvard-Smithsonian Center for Astroph., Cambridge, MA, USA
   \and   Ulugh Beg Astron. Inst., Tashkent, Uzbekistan
   \and   Inst.\ for Astrophys.\ Research, Boston Univ.,\ MA, USA
   \and   Astronomical Institute, Osaka Kyoiku University, Japan
   \and   Astro Space Centre of Lebedev Physical Inst., Moscow, Russia
   \and   Abastumani Astrophysical Observatory, Georgia
   \and   Mets\"ahovi Radio Obs., Helsinki Univ.\ of Technology, Finland
   \and   Korea Astronomy and Space Science Institute, South Korea
   \and   University of Southampton, UK                                            
   \and   Tuorla Observatory, Univ.\ of Turku, Piikki\"{o}, Finland
   \and   Michael Adrian Observatory, Trebur, Germany   
   \and   Cardiff University, Wales, UK                                                      
   \and   Nordic Optical Telescope, Santa Cruz de La Palma, Spain
   \and   Agrupaci\'o Astron\`omica de Sabadell, Spain
   \and   Dept. of Phys., Univ. of Colorado, Denver, USA
   \and   Crimean Astrophysical Observatory, Ukraine   
   \and   Radio Astron.\ Lab.\ of Crimean Astroph.\ Observatory, Ukraine
   \and   ASI Science Data Centre, Frascati, Italy
 \and  INAF, Istituto di Radioastronomia, Sezione di Noto, Italy
 \and  School of Cosmic Physics, Dublin Inst. for Adv. Studies, Ireland
 \and  Cork Institute of Technology, Cork, Ireland
 \and Moscow Univ., Crimean Lab. of Sternberg Astron. Inst., Ukraine
 \and Isaac Newton Institute of Chile, Crimean Branch, Ukraine
 \and Special Astrophysical Observatory, N. Arkhyz, Russia
 \and Instituto de Radioastronom\'{i}a Milim\'{e}trica, Granada, Spain
 \and ARIES, Manora Peak, Nainital, India
 }

   \date{}

 
  \abstract
   {The quasar \object{3C 279} is among the most extreme blazars in terms of luminosity and variability of
flux at all wavebands. Its variations in flux and polarization are quite complex and therefore require
intensive monitoring observations at multiple wavebands to characterise and interpret the observed changes.}
   {In this paper, we present radio-to-optical data taken by the WEBT, supplemented by our VLBA and RXTE
observations, of 3C~279. Our goal is to use this extensive database to draw inferences regarding the physics
of the relativistic jet.}
   {We assemble multifrequency light curves with data from 30 ground-based observatories and the space-based
instruments SWIFT (UVOT) and RXTE, along with linear polarization vs.\ time in the optical $R$ band. In addition,
we present a sequence of 22 images (with polarization vectors) at 43 GHz at resolution 0.15 milliarcsec, obtained
with the VLBA. We analyse the light curves and polarization, as well as the spectral energy distributions at
different epochs, corresponding to different brightness states.}
   {We find that the IR-optical-UV continuum spectrum of the variable component corresponds to a power law
with a constant slope of $-1.6$, while in the 2.4--10
keV X-ray band it varies in slope from $-1.1$ to $-1.6$. The steepest X-ray spectrum
occurs at a flux minimum. During a decline in flux from maximum in late 2006, the optical and 43 GHz
core polarization vectors rotate by $\sim 300$\degr.
}
  {The continuum spectrum agrees with steady injection of relativistic electrons with a power-law energy
distribution of slope $-3.2$ that is steepened to $-4.2$ at high energies by radiative losses. The X-ray emission
at flux minimum comes most likely from a new component that starts in an upstream section of the jet where
inverse Compton scattering of seed photons from outside the jet is important. The rotation of the polarization
vector implies that the jet contains a helical magnetic field that extends $\sim 20$ pc past the 43 GHz
core.}


   \titlerunning{Monitoring of 3C 279 during 2006--2007}
   \authorrunning{Larionov et al.}
   \maketitle
%

\section{Introduction\label{sect:intro}}

Blazars form a class of active galactic nuclei in which the spectral energy distribution is dominated by
highly variable nonthermal emission from relativistic jets that point almost directly along the line of sight.
The quasar-type blazar \object{3C~279}, at redshift $z=0.538$ \citep{Burbidge1965}, is one of the most
intensively studied objects of this class, owing to its pronounced variability of flux across the
electromagnetic spectrum (by more than $5^m$\ in optical bands) and high optical polarization (highest observed
value of 45.5\% in $U$ band, \citealt{Mead1990}). Because of this, 3C~279 has been the target of many
multiwavelength campaigns,
mounted in an effort to learn about the physics of the jet and the high-energy plasma that it contains.
For example, in early 1996 the source was observed at a high $\gamma$-ray state,
with extremely rapid flux variability, by the EGRET detector of the {\it Compton} Gamma Ray
Observatory and at longer 
wavelengths near a historical maximum in brightness  \citep{Wehrle1998}. The $\gamma$-ray
flare was coincident with an X-ray outburst without any lag longer than 1 day. However, while one can draw
some important conclusions from such campaigns, the behaviour of 3C~279 is typically too complex to
characterise and relate conclusively to physical aspects of the jet from short-term light curves at a few
wavelengths.

Ultra-high resolution observations with the Very Long Baseline Array (VLBA) have demonstrated that EGRET-detected 
blazars possess the most highly relativistic jets among compact flat spectrum radio sources 
\citep{Jorstad2001, Kellermann2004}. This is inferred from the appearance of superluminally moving
knots in the jet separating from a bright, compact, stationary ``core'' in the VLBA images.
Apparent speeds in 3C~279 range from 4$c$ to 16$c$ \citep[e.g.,][and references therein]{Jorstad2004}.
From an analysis of both the apparent speeds and the time scales of flux decline of individual knots in 3C~279,
\citet{Jorstad2005} derived a Doppler beaming factor $\delta = 24\pm6$, a Lorentz factor of the jet
flow $\Gamma = 16\pm3$, and an angle between the jet axis and line of sight $\Theta \approx 2^\circ$.
An analysis of the times of high $\gamma$-ray flux and superluminal ejections of $\gamma$-ray blazars indicates a statistical
connection between the two events \citep{Jorstad2001b}. An intensive set of multi-waveband monitoring
and VLBA observations of BL~Lac confirms this connection between high-energy flares and
superluminal knots in at least one blazar \citep{Marscher2008}.

Given the complexity of the time variability of nonthermal emission in blazars, a more complete observational
dataset than customarily obtained is needed to maximize the range of conclusions that can be drawn concerning
jet physics. To this end, \citet{Chatterjee2008} have analysed a decade-long dataset containing
radio (14.5 GHz), single-colour ($R$-band) optical, and X-ray light curves as well as VLBA images at 43 GHz of 3C~279. They
find strong correlations between the X-ray and optical fluxes, with time delays that change from X-ray leading
optical variations to vice-versa, with simultaneous variations in between. Although the radio variations
lag behind those in these wavebands by more than 100 days on average, changes in the jet direction on time scales
of years are correlated with and actually lead long-term variations at X-ray energies. The current paper 
extends these observations
to include multi-colour optical and near-infrared light curves, as well as linear polarization in the optical
range and in the VLBA images, although over a more limited range in time (2006.0-2007.7). The radio-to-optical data presented in this paper have been acquired during a multifrequency campaign organized by the Whole Earth Blazar Telescope (WEBT)\footnote{{\tt http://www.oato.inaf.it/blazars/webt/} \citep[see e.g.][]{Bottcher2005, Villata2006, Raiteri2007}.}  Our dataset includes the data from the 2006 WEBT campaign presented by \citet{Bottcher2007} as well as some of the data analysed
by \citet{Chatterjee2008}. The first of these papers found a spectral hardening during flares 
that appeared delayed with respect to a rising optical flux. The authors interpreted such
behaviour in terms of inefficient particle acceleration at optical bands. \citet{Chatterjee2008} came to a
similar conclusion based on the common occurrence of X-ray (from synchrotron self-Compton emission) variations
leading optical synchrotron variations, with the latter involving higher-energy electrons than the former.

Our paper is organized as follows:
Section~\ref{sect:observ} outlines the procedures that we used to process and analyse the data. Section~\ref{sect:lightcurves} presents and analyses
the multifrequency light curves, Sect.~\ref{sect:kinematics} discusses the kinematics and polarization of the radio jet as
revealed by the VLBA images, and Sect.~\ref{sect:optpol} gives the results of the optical polarimetry.  In Sect.~\ref{sect:radlightcurves} we
derive and discuss the time lags among variations at different frequencies, while in Sect.~\ref{sect:SED} we do the same
for the radio-to-X-ray spectral energy distribution constructed at four different flux states. In Sect.~\ref{sect:discuss} we
discuss the implications of our observational results with respect to the physics of the jet in 3C~279. 

\section{Observations, data reduction and analysis\label{sect:observ}}

Table \ref{obs} contains a list of observatories participating in the WEBT campaign, 
as well as the bands/frequencies at which we acquired the data for this study. Reduction of the data generally
followed standard procedures. In this section, we summarise the observations and methodology.

\subsection{Optical and near-infrared observations}

We have collected optical and near-IR data from 21 telescopes, listed in Table~\ref{obs}. The data
were obtained as instrumental magnitudes of the source and comparison stars in the same field. The finding chart can be found at WEBT campaign WEB-page. We used optical and NIR calibration of standard stars from \citet{Raiteri1998} and \citet{Gonzalez-Perez2001}.

We carefully assembled and ``cleaned'' the optical light curves \citep[see, e.g.,][]{Villata2002}. When necessary,
we applied systematic corrections (mostly caused by effective wavelengths differing from standard $BVR_CI_C$
bandpasses) to the data obtained from some of the participating teams, to match the calibration of the
sources of data that use standard instrumentation and procedures. The resulting offsets do not generally
exceed 0\fm01--0\fm03 in $R$ band.

\subsection{Radio observations}
We collected radio data from 8 radio telescopes, listed in Table~\ref{obs}.
Data binning and cleaning was needed in some cases to reduce the noise. 
We report also VLBA observations performed roughly monthly with the VLBA at 43 GHz ($\lambda$7mm)
in both right and left circular polarizations throughout 2006-2007 (22 epochs). In general, all 10 antennas of the
VLBA were used to reach ultra-high resolution, $\sim$0.15 milliarsecond (mas), for
imaging. However, at a few epochs 1-2 antennas failed due to weather
or technical problems. We calibrated and reduced the data in the same manner 
as described in \citet{Jorstad2005} using the AIPS and Difmap software packages.
The electric vector position angle (EVPA) of linear polarization was obtained by comparison
between the Very Large Array (VLA) and VLBA integrated EVPAs for sources
OJ~287, 1156+295, 3C~273, 3C~279, BL~Lac, and 3C~454.3 that are
common to both our sample and the VLA/VLBA polarization calibration
list\footnote{\tt http://www.vla.nrao.edu/astro/calib/polar}, at epochs simultaneous
within 2-3 days (10 epochs out of 22). At the remaining epochs, we performed
the EVPA calibration using the D-term method \citep{Gomez2002}. We modelled
the calibrated images by components with circular Gaussian brightness distributions, a procedure that allows us
to represent each image by a sequence of components described by flux density,
FWHM angular diameter, position (distance and angle) relative to the core, and polarization
parameters (degree and position angle). 

\subsection{UV observations by Swift}

In 2007, the Swift satellite observed 3C~279 at 15 epochs from January 12 to July 14, covering both bright and faint states of the source. The UVOT instrument \citep{Roming2005} acquired data in the $V$, $B$, $U$, UV$W1$, UV$M2$, and UV$W2$ filters. These data were processed with the {\tt uvotmaghist} task of the HEASOFT package remotely through the Runtask Hera facility\footnote{\tt http://heasarc.gsfc.nasa.gov/hera}, with software and calibration files updated in May 2008. Following the recommendations given by \citet{Poole2008} for UVOT photometry, we extracted the source counts from a circle using a $5''$ aperture radius, and determined the background counts from a surrounding source-free annulus. Since the errors of individual data points are of the order of 0\fm1, we use daily binned data for further analysis.

\subsection{X-ray observations}
We observed 3C~279 with the Rossi X-ray Timing Explorer (RXTE) roughly 3 times per week, with 1--2 ks
exposure during each pointing. This is a portion of a long-term monitoring program
from 1996 until the present \citep{Marscher2006, Chatterjee2008}.
Fluxes and spectral indices over the energy range from 2.4 keV to 10 keV were computed
with the X-ray data analysis software FTOOLS and XSPEC along with the faint-source
background model provided by the RXTE Guest Observer Facility.  We assumed a power-law continuum
spectrum for the source with negligible photoelectric absorption in this energy range.

\begin{table}
\begin{minipage}[t]{\columnwidth}
\caption{Ground-based observatories participating in this work.}
\label{obs}
\centering
\renewcommand{\footnoterule}{}  
\begin{tabular}{l r c}
\hline\hline
Observatory    & Tel.\ diam.     & Bands\\
\hline
\multicolumn{3}{c}{\it Radio}\\
Crimean (RT-22), Ukraine & 22 m          & 36 GHz          \\
Mauna Kea (SMA), USA     &$8 \times 6$ m\footnote{Radio interferometer including 8 antennas of diameter 6 m.}
 & 230, 345 GHz      \\
Medicina, Italy          & 32 m          & 5, 8, 22  GHz        \\
Mets\"ahovi, Finland     & 14 m          & 37 GHz              \\
Noto, Italy              & 32 m          & 43  GHz              \\
Pico Veleta, Spain       & 30 m\footnote{Data obtained with the IRAM 30-meter telescope as a part of standard AGN monitoring program.}
      & 86 GHz              \\
UMRAO, USA               & 26 m          & 5, 8, 14.5 GHz      \\
RATAN-600, Russia & 600 m\footnote{Ring telescope.}
    & 1, 2, 5, 8, 11, 22 GHz      \\
\hline
\multicolumn{3}{c}{\it Near-infrared}\\
Campo Imperatore, Italy  & 110 cm        & $J, H, K$          \\
\hline
\multicolumn{3}{c}{\it Optical}\\
Abastumani, Georgia      &  70 cm        & $R$                \\
Armenzano, Italy         &  40 cm        & $B, V, R, I$       \\
Belogradchik, Bulgaria   &  60 cm        & $R$                \\
Bordeaux, France         &  20 cm        & $V$                \\
Catania, Italy           &  91 cm        & $U, B, V$          \\
COMU Ulupinar, Turkey    &  30 cm        & $V, R, I$          \\
Crimean (AP-7), Ukraine  &  70 cm        & $B, V, R, I$       \\
Crimean (ST-7), Ukraine  &  70 cm        & $B, V, R, I$       \\
Crimean (ST-7, pol.), Ukraine  &  70 cm  & $R$                \\
Lulin, Taiwan            &  40 cm        & $R$                \\
Michael Adrian, Germany  & 120 cm        & $R$                \\
Mt.Maidanak, Uzbekistan  &  60 cm        & $B, V, R, I$       \\
Osaka Kyoiku, Japan      &  51 cm        & $R$                \\
Roque (KVA), Spain       &  35 cm        & $R$                \\
Roque (LT), Spain        & 200 cm        & $R$                \\
Roque (NOT), Spain       & 256 cm        & $U, B, V, R, I$    \\
Sabadell, Spain          &  50 cm        & $R$                \\
Sobaeksan, South Korea   &  61 cm        & $B, V, R, I$       \\
St.\ Petersburg, Russia  &  40 cm        & $B, V, R, I$       \\
Torino, Italy            & 105 cm        & $B, V, R, I$       \\
Valle d'Aosta, Italy     &  81 cm        & $R, I$             \\
Yunnan, China            & 102 cm        & $R$                \\
\hline
\end{tabular}
\end{minipage}
\end{table}

\section{X-ray, optical and near-IR light curves\label{sect:lightcurves}}

\begin{figure}
   \centering
  \resizebox{\hsize}{!}{\includegraphics[clip]{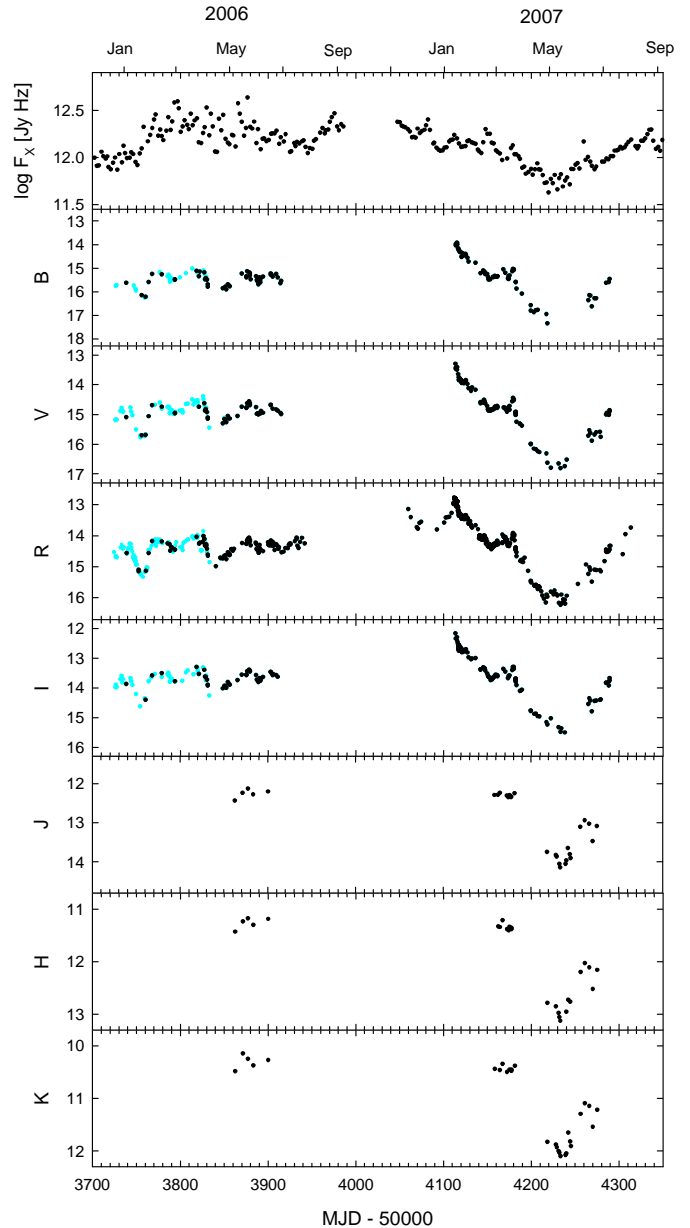}}
      \caption{X-ray, optical and near-IR light curves of 3C 279 in the 2006--2007 observing seasons. Previously published data are marked with blue (light) symbols.}
         \label{bvrijhk}
   \end{figure}

Figure \ref{bvrijhk} displays the X-ray, optical, and near-IR light curves of the 2006--2007 observing seasons. Analysis of the 2006 data is given in a previous WEBT campaign paper \citep{Bottcher2007}; see also \citet{Collmar2007}.


During nearly the entire 2006 season, 3C~279 was at a moderately bright optical level of $R \approx 14\fm5$, with two noticeable downward excursions in January and April (MJD~53740 and 53830). At the beginning of the 2007 season, 3C~279 was at a brighter level, $R=12\fm8$. Subsequently, over a 100-day period, the flux decreased to $R=16\fm2$. Superposed on this declining trend, we observed a mild outburst around MJD~54150--54180, with amplitude $\approx 0\fm7$. The minimum light level at MJD~54230, though not a record for 3C~279, was close to a level at which \cite{Pian1999} were able to detect the contribution of accretion disc in UV part of 3C~279 spectral energy distribution. Unfortunately, we were not able to detect it due to lack of Swift UVOT data close to these dates.

One of the campaign goals was to evaluate the characteristic time scales and amplitudes of intranight variability. We found that, both in high and low optical states, such rapid variability does not exceed 0\fm02~hour$^{-1}$, while night-to-night variations reached 0\fm1--0\fm2. This can be confirmed by visual inspection of Fig.\ref{bvrijhk}, where the general features of the light curves are not masked by noise.

\begin{figure}
\centering
     \resizebox{\hsize}{!}{\includegraphics[clip]{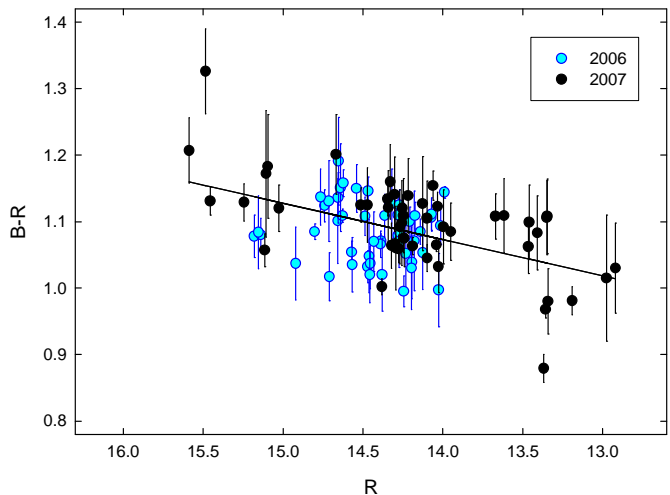}}
      \caption{Colour vs. magnitude dependence of 3C 279; data referring to 2006 and 2007 seasons are marked with different colours.}
         \label{colour}
   \end{figure}

The bluer-when-brighter behaviour is easily seen on Fig.~\ref{colour} (although it is less evident
in the 2006 data). The regression curve is $B-R \propto(0.05\pm0.01)R$.
There are two plausible explanations for the origin of this colour   variation: either we observe simultaneously a constant (or slowly varying) source, e.g., the host galaxy and accretion disc, along with a strongly variable source (in the jet) with a substantially different SED; or, alternatively, the bluer-when-brighter trend is intrinsic to the jet, as found for other blazars, especially BL Lac objects (\citealt{Villata2000,Raiteri2003} for 0716+714; \citealt{Villata2004a,Papadakis2007} for BL Lacertae). In the following we consider the former supposition as better substantiated for 3C~279 behaviour during these observing seasons.

\subsection{Analysis of optical data\label{sect:opt_analys}}

Following the technique developed by Hagen-Thorn \citep[see, e.g.,][and references therein]{Hagen-Thorn2008}, let us suppose that the flux changes within some time interval are due to a single variable
source. If the variability is caused only by its flux\footnote{For the sake of brevity, we use the term ``flux'' instead of the more proper ``flux density.''} variation but the relative SED remains unchanged,
then in the {\it n}-dimensional flux space
$\{F_1, ..., F_n\}$ ({\it n} is the number of spectral bands used in multicolour observations) the observational points must lie on straight lines. The slopes of these lines are the flux ratios for
different pairs of bands as determined by the SED.
With some limitations, the opposite is also true: a linear relation
between observed fluxes at two different wavelengths during some period of flux variability
implies that the slope (flux ratio) does not change. Such a relation for several bands would indicate that
the relative SED of the variable source remains steady and can be derived from the slopes
of the lines.

We use magnitude-to-flux calibration constants for optical ($BVRI$) and NIR ($JHK$) bands from \citet{Mead1990}, and for UVOT ($V$, $B$, $U$, $W1$, $M2$, $W2$) -- from \citet{Poole2008}. The Galactic absorption in the direction of 3C~279 was calculated according to Cardelli's extinction law \citep{Cardelli1989} and $A_V=0\fm095$ \citep{Schlegel1998}.

   \begin{figure}
   \centering
     \resizebox{\hsize}{!}{\includegraphics[clip]{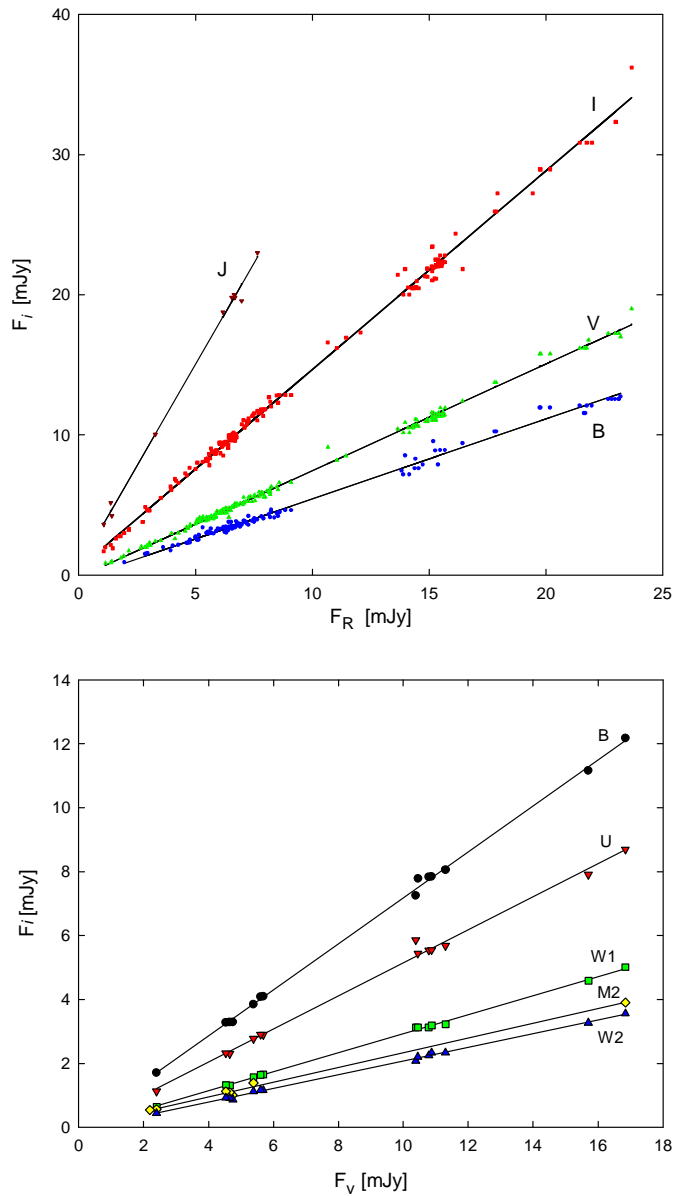}}
      \caption{Flux-flux dependences for optical and NIR (top panel) and UVOT bands (bottom panel).}
         \label{regressions_all}
   \end{figure}

We use the better-sampled $R$-band data to obtain relations in the form of $F_i$ vs. $F_R$. Similarly, in the case of the UVOT data, we derive $F_i$ vs. $F_V$.

Figure~\ref{regressions_all} (upper panel) shows that the method described above holds true: during the entire time covered by our observations, the flux ratios follow linear dependences, $F_i=A_i+B_i\cdot F_R$. Values of $B_i$, along with the slopes of the regressions, can be used to construct the relative SED of the variable source.

A similar analysis can be carried out with UVOT data. Figure~\ref{regressions_all} (bottom panel) demonstrates that the same kind of behaviour is seen in the ultraviolet part of the 3C~279 spectrum. This confirms our supposition about self-similar changes in the SED of the variable source.

\begin{table}
\begin{minipage}[t]{\columnwidth}
\caption{Multicolour properties of the variable optical emission component.}
\label{tab:var}
\centering
\begin{tabular}{c c c c c c }
\hline\hline
Band    & $\log \nu$ (Hz) & $A_\lambda$ & N & r & $\log B$\\
(1) & (2) & (3) & (4) &(5) & (6)\\
\hline
\multicolumn{6}{c}{\it Ground-based observations}\\
\hline
K &14.1307	&0.010	&28	&0.989	&0.8286\\
H &14.2544	&0.016	&25	&0.998	&0.6688\\
J &14.3733	&0.026	&10	&0.998	&0.4609\\
I &14.5740	&0.055	&169	&0.998	&0.1520\\
R &14.6642	&0.076	&	-&-	&0.0000\\
V &14.7447	&0.095	&185	&0.998	&-0.1180\\
B &14.8336	&0.123	&121	&0.996	&-0.2434\\
\hline
\multicolumn{6}{c}{\it UVOT observations}\\
\hline
V &14.7445	&0.093	& -	&-	&-0.1180\\
B &14.8407	&0.123	&14	&0.995	&-0.2607\\
U &14.9329	&0.147	&14 &0.999	&-0.4045\\
W1 &15.0565	&0.195	&13	&0.999	&-0.6438\\
M2 &15.1286	&0.285	& 6	&0.949	&-0.7306\\
W2 &15.1696	&0.271	&14	&0.998	&-0.7856\\
\hline					
\end{tabular}
\end{minipage}
Columns are as follows: (1) - filter of photometry; (2) - logarithm of the effective frequency of the filter; (3) - extinction at the corresponding wavelength; (4) - number of observations in a given band, (quasi)simultaneous with $R$ (for ground-based observations) and $V$ (for UVOT observations); (5) - correlation coefficient; (6) - logarithm of
the contribution of the variable component in the SED at frequency $\nu$
relative to the contribution at $R$ band.
\end{table}

\begin{figure}
   \centering
       \resizebox{\hsize}{!}{\includegraphics[clip]{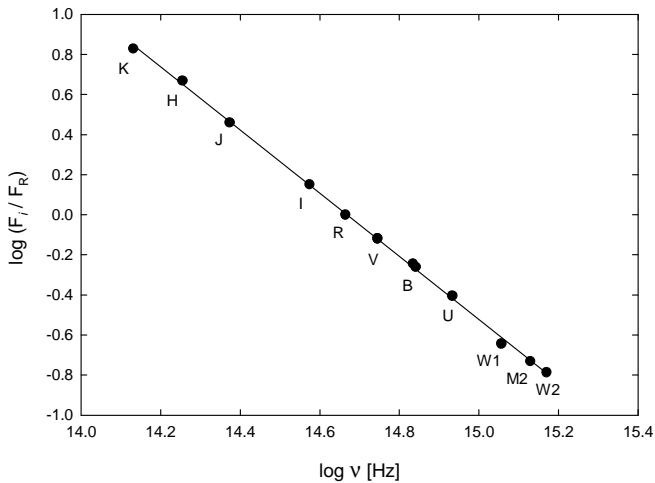}}
      \caption{NIR-optical-UV relative spectral energy distribution of 3C 279 variable source(s), normalized to $R$ band.}
         \label{optical_uvot_sed}
   \end{figure}

Having found the slopes of flux dependences in the near-IR, optical, and UVOT ranges, we are able to construct the relative SED of the variable source in 3C~279. These results are given in Table~\ref{tab:var} (slopes in the UVOT bands are obtained relative to the UVOT $V$ band and fitted to ground-based data) and shown in Fig.~\ref{optical_uvot_sed}.   (The double $B$-band point is caused by different effective wavelengths for the ground-based telescopes and that of UVOT.)
We note that we are unable to judge whether there is only one variable source acting throughout our observations or a number of variable components with the same SED. In any case, in this wavelength range we derive a power-law slope with $\alpha =1.58 \pm 0.01$, in the sense $F_\nu \propto \nu^{-\alpha}$.

   \begin{figure}
   \centering
     \resizebox{\hsize}{!}{\includegraphics[clip]{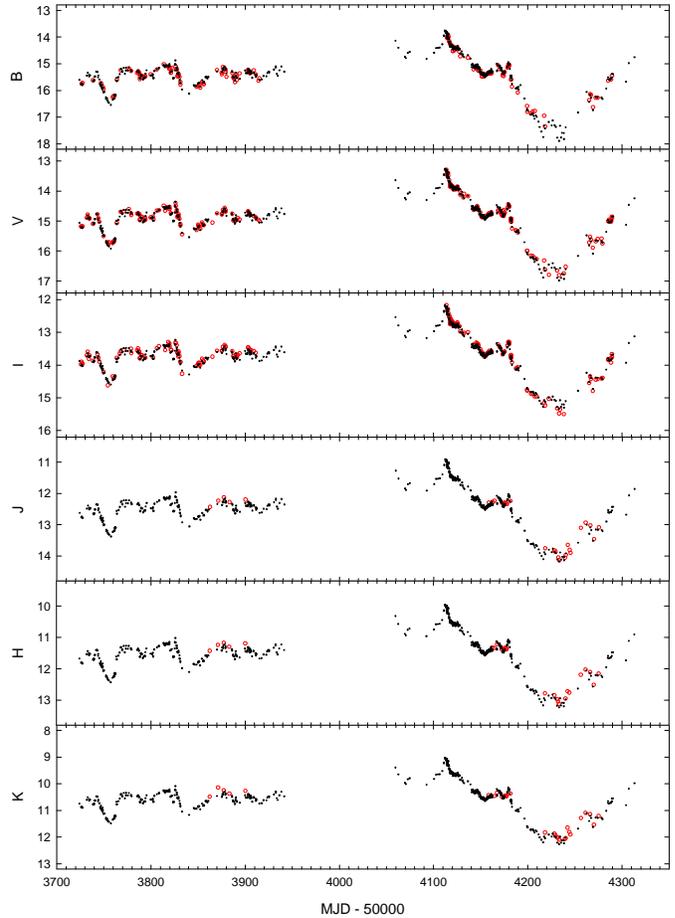}}
      \caption{Comparison of observational data (red circles) and calculated values according to dependences shown in Fig.~\ref{regressions_all} (black dots).}
         \label{fitting_optical2R}
   \end{figure}

It is possible to ``reconstruct'' the light curves in the $BVIJHK$ bands using the dependences shown in Fig.~\ref{regressions_all}. This allows us to (1) check the reliability of this
procedure and (2) form an impression of the variability behaviour without gaps that otherwise are present in less well-sampled bands. 
Figure~\ref{fitting_optical2R} clearly shows that the actual data (red circles) closely match the calculated values (black dots).

Additionally, Figs.~\ref{regressions_all} and~\ref{fitting_optical2R} allow us to conclude that there is no noticeable lag between any of the optical and near-infrared light curves; otherwise, we would see both hysteresis loops in Fig.~\ref{regressions_all}  and misfitting in Fig.~\ref{fitting_optical2R}. If any lags in that wavelength range do exist, they are shorter than a few hours. This contradicts the inference drawn
by \citet{Bottcher2007} based on a more limited dataset.
 
\subsection{X-ray vs. optical fluctuations\label{sect:X_analys}}

From an analysis of the power spectral density of longer-term monitoring observations,
\citet{Chatterjee2008} have concluded that
the X-ray light curve of 3C~279 has a red-noise nature, with fluctuations on short time scales having much lower
amplitude than those on longer time scales. In general, the X-ray variations are either less pronounced or similar
to those at the optical $R$ band. \citet{Chatterjee2008} explain this as the result of the higher energy of electrons
that emit optical synchrotron radiation compared to the wide range of energies -- mostly lower -- of electrons that
scatter IR and optical seed photons to the X-ray band. However, we see the opposite occur between MJD~53770 and 53890,
when the X-ray fluctuations were more dramatic than those at any of the optical bands (see Fig.\ \ref{bvrijhk}).
(Note: the apparent rapid X-ray fluctuations during the minimum at MJD~54210 to 54240 is mostly noise, since the
logarithm of the uncertainty in the fluxes during this time was of order $\pm 0.07$.) We can rule out an
instrumental effect as the cause of the rapid X-ray variability, since the X-ray flux of PKS~1510$-$089,
measured under very similar circumstances, displays only more modest fluctuations. Although a transient, bright,
highly variable X-ray source within $\sim 20'$ of 3C~279 could also have caused the observed fluctuations, no
such source has been reported previously. Hence, we accept the much more likely scenario that the variations
are intrinsic to 3C~279. We discuss a possible physical cause of the rapid X-ray variations in Sect.~\ref{sect:varfluxspec}.

\section{Kinematics and polarization of the radio jet\label{sect:kinematics}}
\begin{figure}
   \centering
\includegraphics[width=8cm,clip]{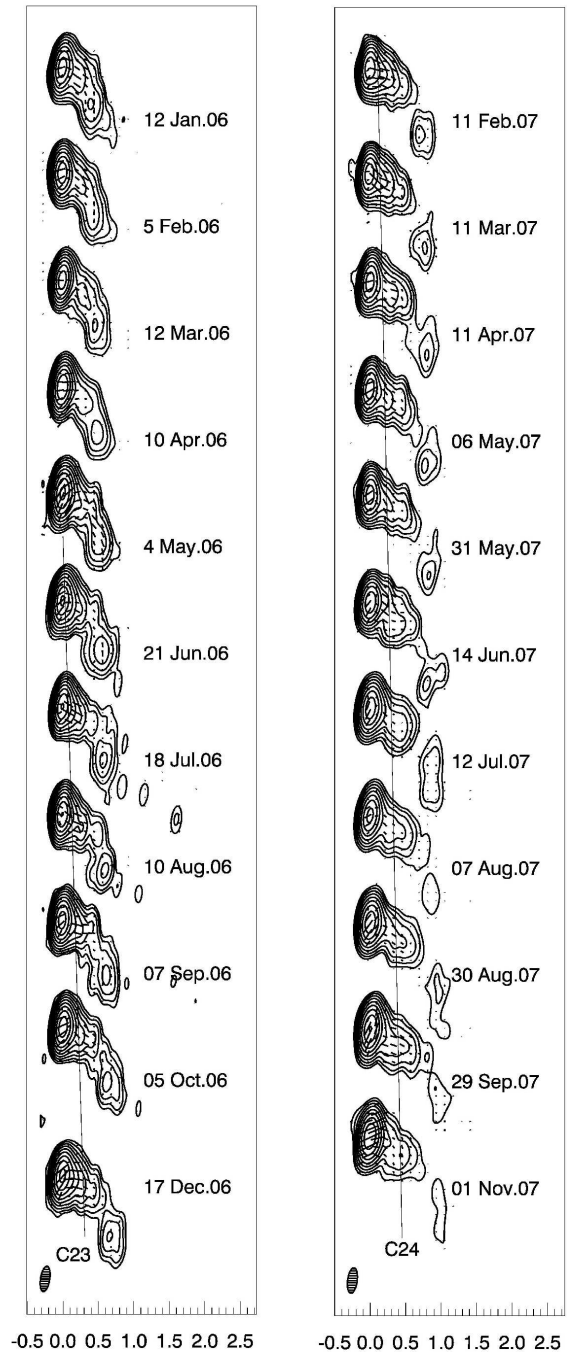}
      \caption{Total intensity VLBA
images of 3C~279 at 43 GHz. Line segments within each image indicate the
direction of polarization electric vectors, with length proportional to the
polarized intensity. The resolution beam is shown by the
cross-hatched ellipse in the lower left corner.  The 
contour levels correspond to 0.25,0.5,1,2,4,8,16,32, and 64\% of the peak intensity
of 17.7~Jy beam$^{-1}$. The solid lines indicate the positions of components
$C23$ and $C24$ that were ejected in 2006.}
         \label{vlba1}
   \end{figure}

The Boston University group observes the quasar 3C~279 with the VLBA at 43~GHz
monthly (if dynamic scheduling works properly) in a program that
started in 2001 \citep[see][]{Chatterjee2008}.
Figure~\ref{vlba1} shows the sequence of total and polarized intensity images of the parsec-scale jet 
obtained with
the VLBA during 2006-2007. The data were reduced and modelled in the same
manner as described in \citet{Jorstad2005}. The images are convolved with the same beam, 0.38$\times$0.14 mas 
at PA=$-9^\circ$, corresponding to the average beam for uniform weighting 
over epochs when all 10 VLBA antennas were in operation (16 epochs out 
of 22 epochs shown). The images reveal motion of two new 
components, $C23$ and $C24$ (we follow the scheme of component designation 
adopted in \citealt{Chatterjee2008}). Figure 7 plots an angular separation 
of the components from the core vs. epoch. Although there is some 
deviation from ballistic motion (especially for $C23$ within 0.3~mas 
of the core), 
a linear dependence fits the data according to the criteria adopted by 
\citet{Jorstad2005}. This gives a high apparent speed for both 
components, $16.5\pm 2.3~c$ and $14.7\pm 0.9~c$, respectively, for $C23$ 
and $C24$ 
(for cosmological parameters $H_\circ=70$ km s$^{-1}$ Mpc$^{-1}$,
$\Omega_m=0.3$, $\Omega_\lambda=0.7$), and yields the following times of 
ejection 
of components (component's coincidence with the core): MJD 53888$\pm$55 
and 54063$\pm$40. In projection on the plane of the sky, these components 
move along similar position angles, $PA\sim -114^\circ$ and $PA\sim 
-120^\circ$. 
The directions are different from the position angles
of components ejected in 2003-2004, $PA\sim -150^\circ$ \citep{Chatterjee2008}, when both the
optical and X-ray activity of the quasar was very modest. The images in Fig.~\ref{vlba1} 
contain another moving component about 1~mas from the core ($C21$, \citealt{Chatterjee2008}), which is a ``relic'' of the southern jet direction 
seen near the core during 2004-2005. Using the method suggested in \citet{Jorstad2005}, 
%
%
we have estimated the variability
Doppler factor, $\delta_{var}$, from the light curves and angular sizes of $C23$ and $C24$,
$\delta_{var}=25$ and $\delta_{var}=29$ respectively, with uncertainties $\sim 20\%$. 
From these values of $\delta_{var}$ and $\beta_{app}$, we
derive the Lorentz factor and viewing angle of the jet, $\Gamma=18\pm 5$,
$\Theta=2.1^\circ\pm 0.3^\circ$ for $C23$, and $\Gamma=18\pm 3$,
$\Theta=1.6^\circ\pm 0.4^\circ$ for $C24$. These values are consistent with those derived
by \citet{Jorstad2005} during the period 1998--2001 when 3C~279 was in a very active
X-ray and optical state, similar to that seen in 2006--07.
 
   \begin{figure}
   \centering
\includegraphics[width=8cm,clip]{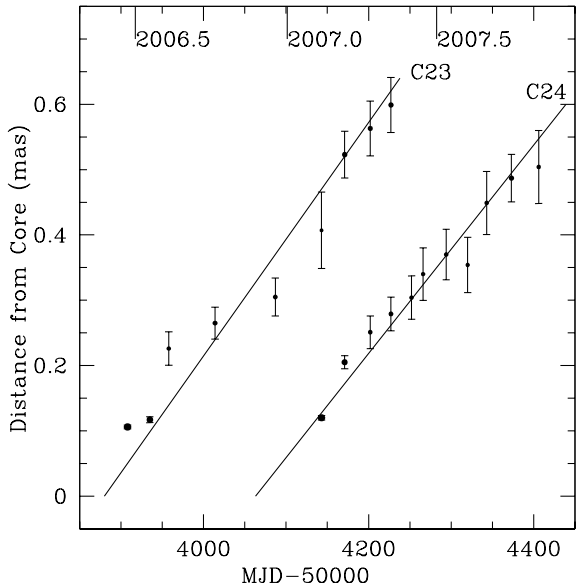}
      \caption{Time evolution of angular separations of $C23$ and $C24$ components.}
         \label{3c279_dist}
   \end{figure}

\begin{figure}
      \centering
     \resizebox{\hsize}{!}{\includegraphics[clip]{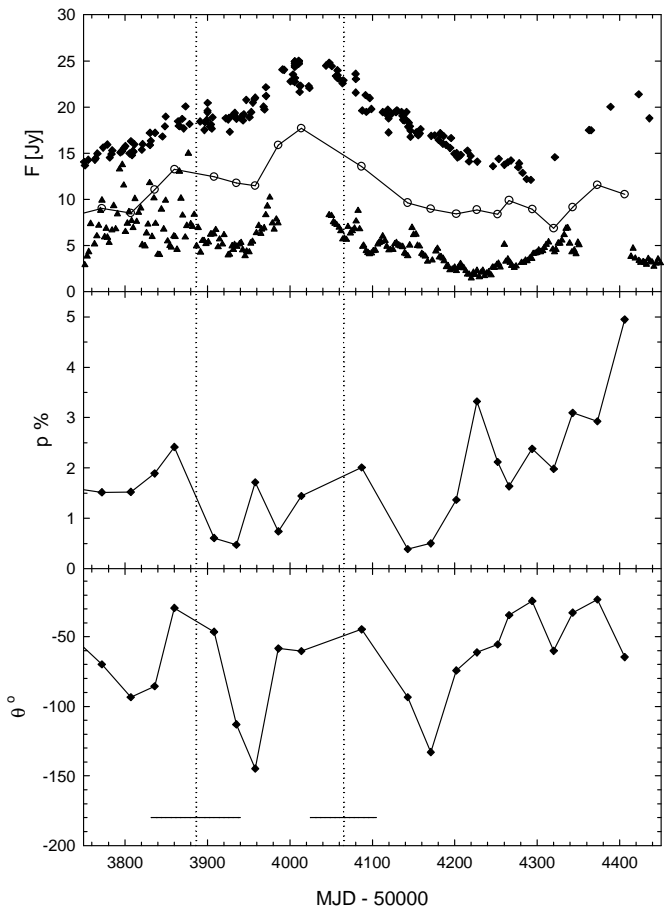}}
      \caption{X-ray and radio observations of 3C~279 during 2006-2007. Solid lines mark VLBI core radiation at
7~mm,  triangles -- X-ray flux (arbitrary units) and diamonds -- 37~GHz flux. Vertical dotted lines
  indicate the times of ejection of components C23 and C24 seen on the VLBA images, and horizontal bars in the bottom panel
-- uncertainty of the times' determination.}
         \label{x-radio}
   \end{figure}

The polarization behaviour in the VLBI core at 7 mm, combined with the jet kinematics discussed
above, helps to define the relationship between the physical state of the jet and the multi-waveband 
variations that we have observed.
Figure~\ref{x-radio} shows the light curve of the VLBI core at 7~mm and also degree and position angle
of polarization in the core. The dotted vertical lines indicate the times of ejection
of components $C23$ and $C24$. The ejections of both components in 2006 occurred near maxima
in the light curve of the VLBI core region, in keeping with the conclusion of
\citet{Savolainen2002} that the emergence of new superluminal knots is responsible for
radio flares observed at $\sim 40$ GHz. Each of the ejections in 2006 coincides with the start
of a rotation of the EVPA in the core of duration $\sim$60-80~days. 
During the rotation the degree of polarization in the core reaches a minimum ($ <0.5$\%). At 
the end of the rotation, the EVPA in the core aligns with the jet direction (see also Fig.~\ref{3c279_polar}). 

Figure~\ref{x-radio} also presents the X-ray and radio (37~GHz) light curves. The ejections of the
components occurred just after maxima in the X-ray light curve. (Although
the peak of the second flare is not covered by our observations due to the proximity of the quasar 
to the Sun, we can infer the approximate time when it occurred by interpolation of the rising and
declining phases of the event.) The X-ray maxima lead the ejections by $\sim$60-80~days. The similarity
of the behaviour at different wavelengths strongly suggests that the two major flares observed
from X-ray to radio wavelengths are associated with disturbances propagating along the jet.


\section{Optical Polarimetry\label{sect:optpol}}

All optical polarimetric data reported here are from the 70cm telescope in Crimea and the 40cm telescope in St.Petersburg, both equipped with nearly identical imaging photometers-polarimeters from St.Petersburg State University. Polarimetric observations were performed using two Savart plates rotated
by 45\degr relative to each another. By swapping the plates, the observer can obtain the relative Stokes $q$ and $u$ parameters from the two split images of each source in the field.
Instrumental polarization was found via stars located near the object under the assumption that their radiation is unpolarized. This is indicated also by the high Galactic latitude (57\degr) and the low level of extinction in the direction of 3C~279 ($A_V=0\fm095$; \citealt{Schlegel1998}).

\begin{figure}
      \centering
     \resizebox{\hsize}{!}{\includegraphics[clip]{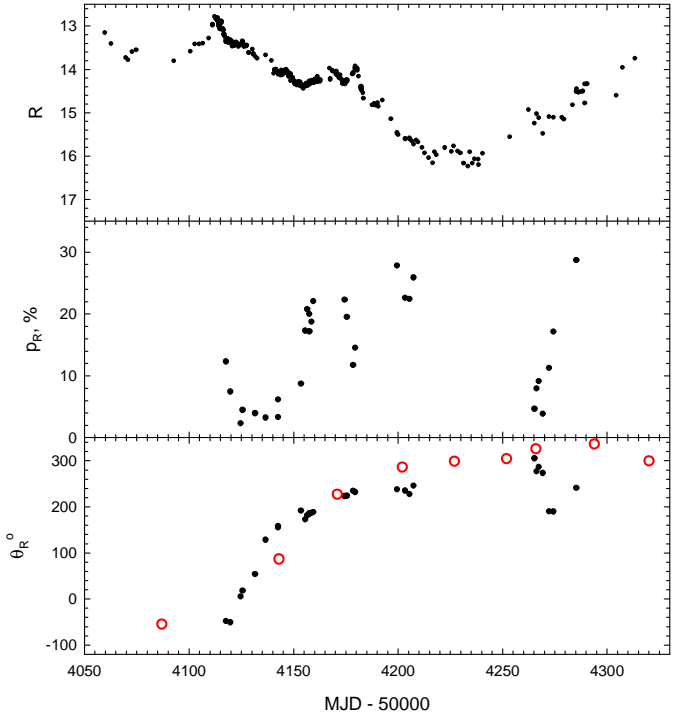}}
      \caption{Optical polarimetric data for 3C~279 obtained during the 2007 campaign. EVPA data for the 43 GHz core
are marked with red (open) symbols in the bottom panel. The $\pm 180$\degr$n$ ambiguity in the EVPAs is resolved by requiring
the change across contiguous epochs to be $<90$\degr. No Faraday rotation correction has been applied to the
43 GHz data; see the text.}
         \label{3c279_polar}
   \end{figure}

The results of polarimetric monitoring of 3C~279 during the 2007 campaign are given in Fig.~\ref{3c279_polar}, with
$R$-band photometry shown in the upper panel for comparison. The most remarkable feature of the polarimetric behaviour is the smooth rotation of the electric-vector position angle (EVPA) of polarization $\theta_R$ from the start of the observing season. This rotation continued, while gradually slowing, over approximately two months. In total, the EVPA rotated by $\sim 300\degr$, eventually stabilizing at $\theta_R \sim $240\degr. Note that, because of the $\pm180$\degr$n$ ambiguity in $\theta_R$, this value is the same as $-120$\degr, the direction of the VLBA inner jet in 3C~279 during 2007 (cf. Fig.~\ref{vlba1}). The bottom panel of Fig.~\ref{3c279_polar} also shows the 7mm radio EVPA (red open circles), which rotates along with the optical EVPA, although with fewer points to define the smoothness of the rotation. The degree of optical polarization $p_R$ varies more erratically than the EVPA, reaching minimum values of $\sim 2\%$ close to the mid-point of rotation of $\theta_R$ and varying between 5\% and 30\% during the remainder of the observational period.

\section{Radio light curves and cross-correlation analysis\label{sect:radlightcurves}}
\begin{figure}
      \centering
     \resizebox{\hsize}{!}{\includegraphics[clip]{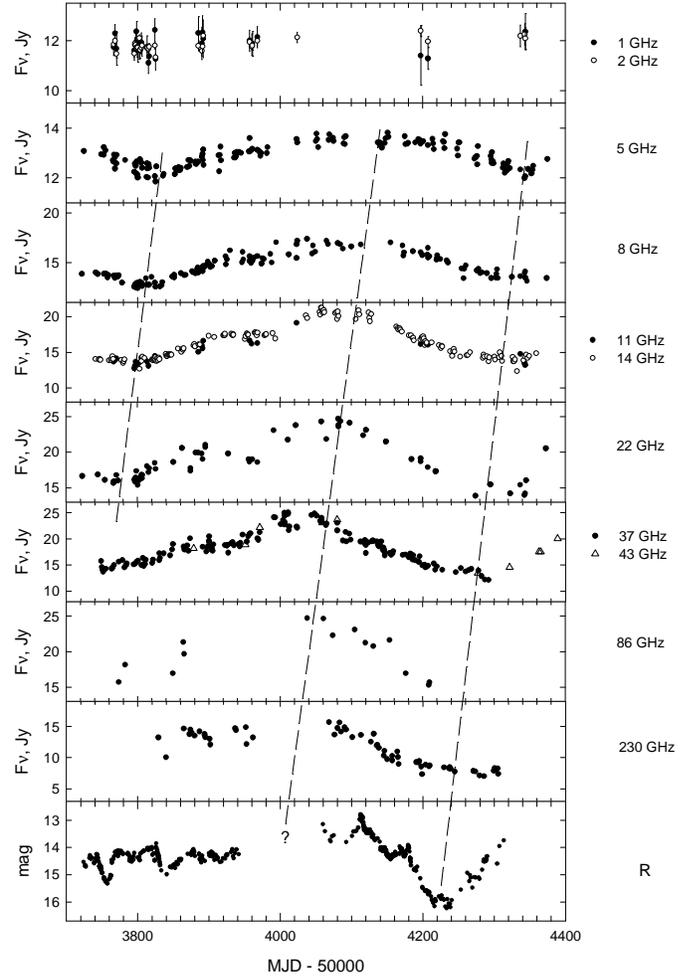}}
      \caption{Radio and $R$-band light curves of 3C 279; dashed slanted lines trace mimima and maxima of
light curves at different frequencies. Error bars are shown only for 1 and 2~GHz data, where they are larger than symbol size.}
         \label{radio}
   \end{figure}

The common way to estimate delays between light curves at different wavelengths is to use the DCF (Discrete Correlation Function; \citealt{Edelson1988, Hufnagel1992}). In the case of inter-optical delays, we have mentioned above that there are no lags between any pairs of wavelengths (see Sect.~\ref{sect:opt_analys}). This conclusion is confirmed with DCF analysis using different pairs of optical bands -- in all the cases the best fit is obtained for 0 days lag. We failed to evaluate DCF lags between optical and radio frequencies for both observational seasons simultaneously, because of the seasonal gap in optical observations caused by solar conjunction and, to a lesser extent, the different time 
sampling. However, inspection by eye of the optical $R$ and radio light curves (Fig.~\ref{radio}) reveals 
progressive lags that increase toward lower frequencies, with a time delay between the minima at $R$ band and 37 GHz in
mid-2007 of $\sim 60$ days. The dashed slanted lines in Fig.~\ref{radio} 
connect the positions of two minima and one maximum that are easily distinguished in the light curves. The question mark in the $R$-band panel denotes the position of a tentative optical maximum, missed due to the seasonal gap.  Such behaviour is not unexpected and confirms the results obtained during earlier
epochs \citep[e.g.,][]{Tornikoski1994,Lindfors2006}. 

\begin{figure}
   \centering
     \resizebox{\hsize}{!}{\includegraphics[clip]{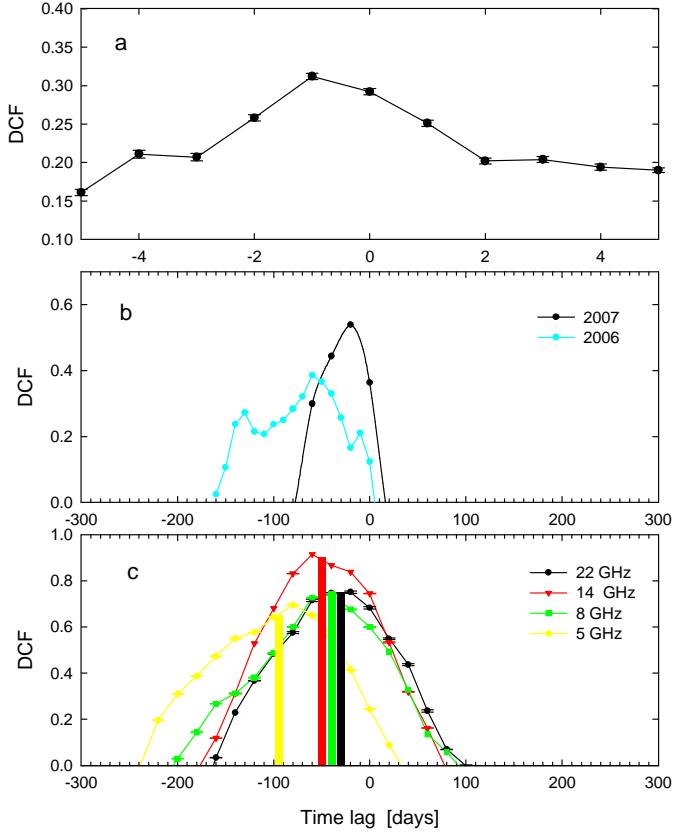}}
      \caption{Discrete correlation functions (a) for optical--X-ray, (b) optical--radio (37GHz) and (c) radio frequencies relative to the 37~GHz light curve. Negative time lag corresponds to lower-frequency radiation changing later than at higher frequencies. Vertical bars in panel (c) indicate the lags corresponding to the DCF centroids. Note that in panel (b), due to seasonal gap in optical range, the DCF is calculated separately for the 2006 and 2007 seasons.}
         \label{DCF}
   \end{figure}

We calculated the optical--radio (37~GHz) delays separately for two observing seasons (Fig.~\ref{DCF}b). The negative values of delays agree with our conclusion made from eye inspection of Fig.~\ref{radio}.
In order to evaluate time lags quantitatively at radio frequencies, we have computed the DCF for the 22~GHz, 14~GHz, 8~GHz and 5~GHz light curves relative to 37~GHz. The results are plotted in Fig.~\ref{DCF}c, where vertical bars indicate the lags corresponding to the DCF centroids. Despite the large uncertainties, this plot confirms our conclusion about variations at lower frequencies lagging behind those at high frequencies. Similar results were found by \citet{Raiteri2003} for \object{S5~0716+714}, \citet{Villata2004b} for \object{BL~Lac}, \citet{Raiteri2005} for \object{AO~0235+164}, \citet{Villata2007} for \object{3C~454.3}.

The upper panel of Fig.~\ref{DCF} displays the X-ray--optical DCF. The correlation is rather weak, with 
changes in the optical ($R$) flux lagging behind X-ray variations by $\sim 1$ day. Inspection by eye suggests
that the correlation is strong during 2007 and weak or non-existent in 2006. The most salient feature in the
light curves, a deep minimum between MJD~54220
and 54250, occurs essentially simultaneously at both X-ray and optical wavebands. As is discussed
by \citet{Chatterjee2008}, the X-ray--optical
correlation fluctuated slowly over the years from 1996 to 2007, both in strength and in the length and
sense of the time lag, with the weakest correlation occurring when the lag is near zero. The latter
behaviour matches that observed in 2006--07. The controlling factor may be slight changes in the jet 
direction over time, which modulates the Doppler factor, light-travel delay across the source, and
the position in the jet where the main observed emission originates \citep{Chatterjee2008}.

\section{Spectral energy distribution\label{sect:SED}}

\begin{figure}
\centering       
\resizebox{\hsize}{!}{\includegraphics[clip]{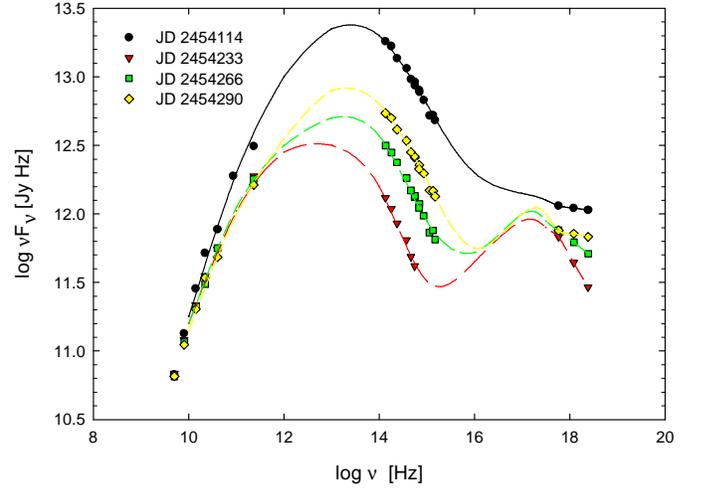}}
      \caption{Radio-to-X-ray spectral energy distributions at various epochs in 2007. For each epoch,
a cubic spline is drawn through the points. The errors of slope in the X-ray part of SED are of the order of 0.1.}
         \label{all_sed}
   \end{figure}

Figure \ref{all_sed} displays the spectral energy distribution (SED) of 3C~279 at four different flux levels
at epochs when we have sufficient data. Note that the non-variable optical component has not been subtracted
in these plots. The overall increase in flux across the entire sampled frequency range strongly suggests that
the nonthermal emission at radio through X-ray wavelengths has a common origin, which we identify with the jet
based on the appearance of new superluminal knots associated with each of the two major outbursts in flux.
The other striking feature of the SEDs is the steepening of the X-ray spectrum at lower flux levels. (Note:
the X-ray spectral index is derived from an exponential smoothing of the values obtained from the
observations, with a smoothing time of 10 days. This is necessary because of the large uncertainties in
individual spectral indices derived when the flux is low.) It is perhaps significant that the highest X-ray
spectral index of 1.6 matches that of the variable optical component, while the flattest X-ray spectrum has
an index that is lower by 0.5. As we discuss below, this has important implications for the energy gains
and losses experienced by the radiating electrons.

\section{Discussion\label{sect:discuss}}

\subsection{Variability in flux and continuum spectrum\label{sect:varfluxspec}}

During the two observing seasons reported here, 3C~279 exhibited flux variations from radio to X-ray wavelengths
(see Fig.\ \ref{bvrijhk}). We observed the shortest time scales of variability,
$\Delta t/\ln (F_{\rm max}/F_{\rm min}) \sim$ days, in the X-ray and optical bands. There were two deep
minima in the X-ray light curve, the first (at the start of 2006) coinciding with a low optical state with at
least two closely spaced minima, and the second essentially simultaneous with a deep optical minimum between
MJD~54210 and 54240. The discrete cross-correlation function indicates that the time lag between the X-ray
and optical bands is essentially zero.

The SEDs in Fig.~\ref{all_sed} demonstrate that the X-ray emission is not simply a continuation of the
optical synchrotron spectrum. This agrees with the conclusion of \citet{Chatterjee2008} that the X-rays
result from synchrotron self-Compton (SSC) scattering. During the highest flux state, the X-ray spectral index
was $\alpha_x = 1.1$, while the optical spectral index of the variable component was $\alpha_{\rm opt} = 1.6$
throughout both observing seasons
(see Fig.\ \ref{optical_uvot_sed}). This can be accommodated within a standard nonthermal source model if
there is continuous injection of relativistic electrons with a power-law energy distribution of slope
$-3.2$ that is steepened to $-4.2$ at high energies by radiative energy losses. This implies that there
is a flatter spectrum at mid- and far-IR wavelengths, below $\sim 10^{14}$ Hz,
with $\alpha_{\rm IR} = 1.1$ that matches that of the high flux state.
At minimum flux, the X-ray spectrum steepens to $\alpha_x = 1.6 = \alpha_{\rm opt}$.

The steep X-ray spectrum during the flux minimum of 2007 implies that the entire synchrotron spectrum above the
spectral turnover frequency had a steep spectral index of 1.6. In the context of the standard nonthermal
source model mentioned above, essentially all of the electrons suffered significant radiative energy losses at
the time of the low state. That is, their radiative loss time scale was shorter than the time required
for them to escape from the emission region. For this to be the case, the injected energy distribution must have
a low-energy cutoff that is far above the rest-mass energy. We can match the time scale of optical and
X-ray variability of 3C~279 if we adopt a magnetic field strength in the emission region of $B \sim 1$~G.
Electrons radiating at the UV wavelengths must then have a Lorentz factor $\gamma \equiv (E/mc2) \sim 6000 B^{-1/2}(\delta/25)^{-1/2}$,
those radiating at a break frequency of $\sim 10^{14}$ Hz have $\gamma \sim 1500 B^{-1/2}(\delta/25)^{-1/2}$,
and those emitting at $10^{12}$ Hz have $\gamma \sim 150 B^{-1/2}(\delta/25)^{-1/2}$. Given the turnover in the SED at $\sim 10^{12}$ Hz
(cf. Fig.\ \ref{all_sed}), we assume that either the variable emitting component becomes self-absorbed near
this frequency or that there is a lower-energy cutoff (or sharp break) in the injected electron energy
distribution at $\gamma \sim 150 B^{-1/2}(\delta/25)^{-1/2}$.
The time scale for energy losses from synchrotron radiation, as measured in our reference frame, is $\sim 4 B^{-3/2}(\delta/25)^{-1/2}$ days
at $\sim 10^{12}$Hz if we adopt a Doppler factor of 25 (see Sect.~\ref{sect:kinematics} above). This is similar to
the minimum X-ray variability time scale that we observed. The time scale for synchrotron energy losses allows
the variability at optical bands to be as fast as $\sim 3$ hours as long as the emission region is smaller than
$\sim 5\times 10^{15} (\delta/25)$ cm.

The 2.4--10 keV X-ray emission results from a range of seed photons scattered by electrons having a range of
energies \citep{McHardy1999}. If there is a break in the spectrum, as we infer, then the bulk of the X-ray
emission is from seed photons at frequencies below the break -- $\nu \sim 10^{12}$--$10^{14}$ Hz --
scattered by electrons with energies below the break -- $\gamma \sim$ 150--1500. The rapid optical variability
in the low flux state implies that the magnetic field remained of similar strength as for the high state.
However, the break frequency must have decreased to nearly $10^{12}$Hz for the SSC X-ray spectrum to have
steepened to $\alpha_x = \alpha_{\rm opt} = 1.6$. This could occur simply from adiabatic cooling if the
emission region expanded by a factor of 10 between the high and low states. However, in this case the magnetic
field would have dropped, increasing the time scale of variability and decreasing the flux
by more than the observed factor of $\sim 10$ \citep[cf., e.g.,][]{MG85}. Alternatively,
the time for electrons to escape the emission region could have increased as the flux declined. For example,
in the shock model of \citet{MG85}, the electrons are accelerated at the shock front and advect toward
the rear of the shocked region until they encounter a rarefaction at a distance $x$ behind the shock. If the
primary energy loss is due to synchrotron radiation, the break frequency $\nu_b \propto x^{-2}B^{-3}$. A factor of 10
increase in the extent (in the direction parallel to the jet axis) of the shocked region would then cause
the break frequency to decrease from $\sim 10^{14}$ to $10^{12}$ Hz. However, this would require a high relative
velocity between the shock front and the rear of the shocked region, implying that the emission near the latter
is relatively poorly beamed and therefore not prominent.

Another possibility, which seems less contrived, is that the
X-ray emission during the flux minimum is not the decayed remnant of the previous event, but rather comes
from the new component that caused the outburst seen during the
last $\sim 100$ days of the 2007 observing season. This component would first be seen well upstream of
the location where the flux reaches a maximum, and therefore the external photon field from the broad
emission-line region and dusty torus would initially be higher than at later times \citep[see][]{Sokolov2005}.
In this case, the break frequency would increase with time as the component
moves downstream and the level of inverse Compton energy losses decreases. This scenario predicts that
there should be a $\gamma$-ray flare from inverse Compton scattering of seed photons originating outside
the jet prior to the main outburst. Although the $\gamma$-ray observations needed to test this were not
undertaken in early May 2007 (the newly-launched AGILE telescope was in the testing phase), future observations
involving GLAST and AGILE will determine whether such precursor flares actually occur.

The higher amplitude of variability at X-ray energies relative to the optical bands between MJD~53770 and 53890
could be the result of the higher sensitivity of the SSC vs.\ the synchrotron flux to changes in the number of
relativistic electrons (quadratic vs.\ linear dependence). Although, as discussed above, the X-rays during high
flux states are scattered seed photons that originally had a range of frequencies from $\sim 10^{12}$ to
$10^{14}$ Hz, the longest time scale of variability -- a few days at $10^{12}$ Hz -- is short enough to be
consistent with the observations. Alternatively, variable optical depths within the emitting region could
cause the density of SSC seed photons at frequencies $\sim 10^{12}$ Hz to
fluctuate without changing the observed synchrotron radiation much,
thereby causing the SSC X-rays to be more highly variable than the optical
synchrotron radiation.

The continuum spectrum from UV to near-IR wavelengths is well described by a power law with constant spectral index
$\alpha_{\rm opt}=1.6$, even as the flux varied by over an order of magnitude. As discussed above, this agrees with
constant injection of relativistic electrons with a power-law energy distribution of slope $-3.2$ that is steepened
to $-4.2$ at high energies by radiative energy losses. Although such steep slopes are not in conflict with
particle acceleration
models, neither are they predicted. Yet the constancy of this steep slope during the campaign implies that the
physical parameters governing the particle acceleration process can be maintained over long periods of time and
from one event in the jet to the next. This presents a challenge to theories for particle acceleration.

The radio light curves from 230 to 1~GHz become progressively smoother with decreasing frequency. As can be
inferred from Fig.~\ref{radio}, there is also a lag in the timing of maxima and minima in the light curves toward
lower frequencies, as well as a delay of 100--150 days between optical variations and those at 5~GHz. This is readily
explained as a consequence of opacity, with an outburst delayed at radio frequencies until the disturbance
reaches the location in the jet where the optical depth is less than unity \citep[see, e.g.,][]{Hughes85,MG85}.
In addition, the radiative lifetime of electrons in the radio-emitting regions is longer than is the case farther
upstream, owing to the lower magnetic field and photon energy density. This prolongs the duration of an outburst --
and therefore the timing of the minima in the light curves -- at lower frequencies. Because of the absence
of optical and 230 GHz data near the global peak in 2006-07, we cannot determine whether there is a significant
delay between the optical and 230 GHz light curves. However, the fact that at the end of the campaign period we see a brightening in the optical, but not in the mm light curve, suggests that the two emission regions are not co-spatial.

\subsection{Polarimetric behaviour}

The main feature of interest in the polarimetric data is the rotation of the $R$-band and 43 GHz core EVPAs between
MJD~54120 and 54200 (see Fig.\ \ref{3c279_polar}). The optical polarization was several percent or less during
the middle of this rotation, but it
was as high as 23\% near the beginning and end. Apparent rotations by $\sim 300$\degr or more can occur
simply from random walks caused by turbulent cells with random magnetic field orientation passing through the
emission region \citep{Jones1988}. In this case, however, the apparent rotation has an extremely low probability
to be as smooth as we have observed \citep{DArcangelo2007}. We therefore conclude that the rotation is intrinsic
to the jet. The coincident rotation of the 43 GHz core and optical polarization vector suggests a common
emission site at the two wavebands. [Note: the discrepancy between 43 GHz and optical EVPAs measured at essentially
the same time could be due to Faraday rotation, which is of the order of tens of degrees and might vary with
location \citep{ZT04,Jorstad2007}.]

The observed behaviour is qualitatively similar to the models suggested by
\citet{Kikuchi1988,Sillanpaa1993,Marscher2008}, where a shock wave or other compressive feature propagating down
the jet traces a spiral path and cycles through the orientations of an underlying helical magnetic field. This
manifests itself through rotation of the position angle of linear polarization as the feature moves outward.
The polarization is low during the rotation because the symmetry of the toroidal component of the helical
magnetic field produces a cancellation of the net linear polarization when integrated over the entire
structure of the feature. On the other hand, the cancellation is only partial, which implies that the
feature does not extend over the entire cross-section of the jet.

Several documented events of optical position angle rotation have  been reported
by \citet{Sillanpaa1993} and \citet{Marscher2008} (both -- \object{BL~Lac}), \citet{Kikuchi1988} (\object{OJ287}),
and \citet{Larionov2008} (\object{S5~0716+71}). The time scale of rotation of the polarization vector in 3C~279
is much longer than in any of these previous cases -- almost two months as compared to $\sim 1$ week. This
difference can be explained by the substantially larger scale of the core of the 3C~279 jet as well as the
greater distance from the central engine \citep[as determined by the intrinsic opening angle of the jet of
$0.4$\degr$\pm 0.2$\degr;][]{Jorstad2005}. The time required for the disturbance to reach and pass through
the millimeter-wave core is therefore longer than in the other blazars.

The first superluminal knot that emerged during our campaign (see Fig.\ \ref{x-radio}) was located at the
centroid of the 43 GHz core on MJD~53888$\pm$55. (It is interesting that this coincided with the end of the
interval of rapid
fluctuations of the optical and, especially, X-ray flux.) The flux subsequently rose at all wavebands, reaching
a peak $\sim 130$ days later. We note that the EVPA of the core at 43 GHz changed by $\sim 130$\degr\ as the flux
rose, although the data are too sparse to
determine whether this was a smooth rotation. The next knot coincided with the core on MJD~54063$\pm$40, $\sim 50$
days after the radio and X-ray flux peak but at a time of multiple, major optical flares. This was
shortly before the start of the optical EVPA rotation featured in Fig.~\ref{3c279_polar}. The timing
suggests that the optical flares and EVPA rotation in early 2007 took place in this knot as it moved
downstream of the core. This contrasts with the case of BL~Lac, in which
the region with helical magnetic field lies upstream of the millimeter-wave core. This led \citet{Marscher2008}
to associate the helical field with the acceleration and collimation zone (ACZ) of the jet flow, downstream of which
the flow is turbulent. For the polarization rotation to occur downstream of the 43 GHz core, either the core lies
within the ACZ in 3C~279 or the helical field can persist beyond the ACZ. The latter possibility is favoured
by \citet{Gabuzda2008} based on circular polarization observations of blazars. The downward curvature of the
EVPA vs.\ time curve of Fig.\ \ref{3c279_polar} implies that the pitch angle of the helix becomes smaller -- i.e.,
the helix opens up -- with distance from the core. The time scale for this to occur $\sim 60$ days, which
corresponds to a distance of $\sim 20$ pc along the jet of 3C~279, given the kinematics and angle to the line
of sight of the jet derived in Sect.~{\ref{sect:kinematics}}.

\section{Conclusions\label{sect:conclusions}}

By following the evolution of the flux at radio, near-IR, optical, UV, and X-ray frequencies and the linear
polarization at radio and optical bands intensely over a two-year period, we have uncovered patterns that
reveal key aspects of the physics in the relativistic jet in 3C~279. The IR-optical-UV continuum spectrum
of the variable component follows a power law with a constant slope of $-1.6$, while that in the 2.4--10
keV X-ray band varies in slope from $-1.1$ to $-1.6$. This agrees with the expectations of an emission
region into which electrons are steadily injected with a power-law energy distribution of slope $-3.2$ that
is modified to a slope of $-4.2$ at high energies owing to radiative losses. The steepest X-ray spectrum
occurs at a flux minimum. The least contrived explanation is that the X-ray emission at this time comes
from a new component in an upstream section of the jet where the radiative losses from inverse Compton
scattering of seed photons from the broad emission-line region are important. If this is the case, then
a $\gamma$-ray flare should precede the rising portion of a multi-waveband outburst, a prediction that
can be tested with GLAST and AGILE along with intensive multi-waveband monitoring.

During the decline of flux from the maximum in early 2007, we observe a rotation of the optical and 43 GHz
core polarization vectors totaling $\sim 300$\degr. The smoothness of the rotation leads us to conclude that,
as in BL~Lac \citep{Marscher2008} and possibly other blazars, the jet contains a helical magnetic field.
However, in contrast with BL~Lac, the region of helical field in 3C~279 extends $\sim 20$ pc past the 43 GHz
core. Given the paucity of well-sampled optical polarization monitoring over periods of time longer than
1--2 weeks, such rotations may be the norm in blazars. If so, more extensive future polarization monitoring
should uncover many more examples, allowing more general inferences to be drawn regarding the magnetic field
structure in jets and its relationship to the dynamics of the flow.

\begin{acknowledgements}
The research at Boston University was funded in part by the National Science Foundation 
through grant AST-0406865 and by NASA through RXTE Guest Investigator grant NNX06AG86G, and
Astrophysical Data Analysis Program grant NNX08AJ64G. The VLBA is an instrument
of the National Radio Astronomy Observatory, a facility of the National Science Foundation, USA,
operated under cooperative agreement by Associated Universities, Inc.
This work is partly based on observations made with the Nordic Optical Telescope, operated
on the island of La Palma jointly by Denmark, Finland, Iceland,
Norway, and Sweden, in the Spanish Observatorio del Roque de los
Muchachos of the Instituto de Astrofisica de Canarias. Partly based on observations with the Medicina and Noto telescopes operated by INAF - Istituto di Radioastronomia.
This research has made use of data from the University of Michigan Radio Astronomy Observatory,
which is supported by the National Science Foundation and by funds from the University of Michigan. The Submillimeter Array is a joint project between the Smithsonian
Astrophysical Observatory and the Academia Sinica Institute of Astronomy and
Astrophysics and is funded by the Smithsonian Institution and the Academia
Sinica.
The Liverpool Telescope is operated on the island of La Palma by Liverpool John Moores University in the Spanish Observatorio del Roque de los Muchachos of the Instituto de Astrofisica de Canarias with financial support from the UK Science and Technology Facilities Council.
The Mets\"ahovi team acknowledges the support from the Academy of Finland.
AZT-24 observations are made within an agreement between  Pulkovo,
Rome and Teramo observatories.
The Torino team acknowledges financial support by the Italian Space Agency through contract 
ASI-INAF I/088/06/0 for the Study of High-Energy Astrophysics. Y. Y. Kovalev is a Research Fellow of the Alexander von Humboldt Foundation. \mbox{RATAN--600} observations are partly supported by the Russian Foundation for Basic Research (projects 01-02-16812, 05-02-17377, 08-02-00545). We thank Tuomas Savolainen for useful discussion. This paper is partly based on observations carried out at the 30-m telescope of IRAM, which is supported by INSU/CNRS (France), MPG (Germany) and IGN (Spain). I.A. acknowledges support by the CSIC through an I3P contract, and by the ``Ministerio de Ciencia e Innovaci\'on" and the European Fund for Regional Development through grant AYA2007-67627-C03-03. ACG's and WY's work is supported by NNSF of China grant No. 10533050.
\end{acknowledgements}


\begin{thebibliography}{50}
\bibitem[B{\"o}ttcher et al.(2005)]{Bottcher2005} B{\"o}ttcher et al., 2005, \apj, 631, 169

\bibitem[B{\"o}ttcher et al.(2007)]{Bottcher2007} B{\"o}ttcher, M., Basu, 
S., Joshi, M., et al., 2007, \apj, 670, 968

\bibitem[Burbidge 
\& Rosenberg(1965)]{Burbidge1965} Burbidge, E.~M., \& Rosenberg, F.~D., 1965, \apj, 142, 1673 

\bibitem[Cardelli, Clayton, 
\& Mathis(1989)]{Cardelli1989} Cardelli, J.~A., Clayton, G.~C., \& Mathis, J.~S., 1989, \apj, 345, 245 


\bibitem[Chatterjee et al.(2008)]{Chatterjee2008} Chatterjee, R., Jorstad, S.~G., Marscher, A.~P., et al., 2008, \apj, 689, in press; {\tt http://arxiv.org/abs/0808.2194}

\bibitem[Collmar et al.(2007)]{Collmar2007} Collmar, W., B{\"o}ttcher, M., 
Krichbaum, T., et al., 2007, ArXiv e-prints, 710, {\tt http://arxiv.org/abs/0710.1096} 

\bibitem[D'Arcangelo et al.(2007)]{DArcangelo2007} D'Arcangelo, F.~D., Marscher, A.~P., Jorstad, S.~G., et al., \apjl, 659, L107

\bibitem[Edelson 
\& Krolik(1988)]{Edelson1988} Edelson, R.~A., \& Krolik, J.~H., 1988, \apj, 333, 646 

\bibitem[Gabuzda et al.(2008)]{Gabuzda2008} Gabuzda, D.~C., Vitrishchak, V.~M., Mahmud, M., O'Sullivan, S.~P.,
2008, \mnras, 384, 1003

\bibitem[G\'omez et al.(2002)]{Gomez2002} G\'omez J.~L., Marscher, A.~P., Alberdi, A., et al., 2002,
VLBA Scientific Memo 30 (Socorro:NRAO)

\bibitem[Gonz{\'a}lez-P{\'e}rez, Kidger, 
\& Mart{\'{\i}}n-Luis(2001)]{Gonzalez-Perez2001} Gonz{\'a}lez-P{\'e}rez, J.~N., Kidger, M.~R., \& Mart{\'{\i}}n-Luis, F., 2001, \aj, 122, 2055 

\bibitem[Hagen-Thorn et al.(2008)]{Hagen-Thorn2008} Hagen-Thorn, V.~A., 
Larionov, V.~M., Jorstad, S.~G., et al., 2008, \apj, 672, 40

\bibitem[Hufnagel 
\& Bregman(1992)]{Hufnagel1992} Hufnagel, B.~R., \& Bregman, J.~N., 1992, \apj, 386, 473 

\bibitem[Hughes, Aller, \& Aller(1985)]{Hughes85} Hughes, P.~A., Aller, H.~D., \& Aller,M~F., 1985, \apj, 298, 301

\bibitem[Jones(1988)]{Jones1988} Jones, T.~W., 1988, \apj, 332, 678

\bibitem[Jorstad et al.(2001a)]{Jorstad2001} Jorstad, S.~G., Marscher, 
A.~P., Mattox, J.~R., et al., 2001a, \apjs, 134, 181 

\bibitem[Jorstad et al.(2001b)]{Jorstad2001b} Jorstad, S.~G., Marscher, 
A.~P., Mattox, J.~R., et al., 2001b, \apj, 556, 738

\bibitem[Jorstad et al.(2004)]{Jorstad2004} Jorstad, S.~G., Marscher, 
A.~P., Lister, M.~L., et al., 2004, \aj, 127, 3115 

\bibitem[Jorstad et al.(2005)]{Jorstad2005} Jorstad, S.~G., Marscher, 
A.~P., Lister, M.~L., et al., 2005, \aj, 130, 1418 

\bibitem[Jorstad et al.(2007)]{Jorstad2007} Jorstad, S.~G., Marscher, 
A.~P., Stevens, J.~A., et al., 2007, \aj, 134, 799 

\bibitem[Kellermann et al.(2004)]{Kellermann2004} Kellermann, K.~I., 
Lister, M.~L., Homan, D.~C., et al., 2004, \apj, 609, 539 

\bibitem[Kikuchi et al.(1988)]{Kikuchi1988} Kikuchi, S., Mikami, Y., Inoue, 
M., Tabara, H., \& Kato, T., 1988, \aap, 190, L8 

\bibitem[Larionov et al.(2008)]{Larionov2008} Larionov, V., Konstantinova, T., Kopatskaya, E., et al., 
2008, The Astronomer's Telegram, 1502

\bibitem[Lindfors et al.(2006)]{Lindfors2006}Lindfors, E.~J., T\"urler, M., Valtaoja, E., et al. 2006, \aap, 456, 895

\bibitem[Marscher(2006)]{Marscher2006} Marscher, A.~P., 2006, Blazar 
Variability Workshop II: Entering the GLAST Era, 350, 155 

\bibitem[Marscher et al.(2008)]{Marscher2008} Marscher, A.~P., Jorstad, 
S.~G., D'Arcangelo, F.~D., et al., 2008, \nat, 452, 966 

\bibitem[Marscher \& Gear(1985)]{MG85} Marscher, A.~P., \& Gear, W.~K., 1985, \apj, 298, 114

\bibitem[McHardy et al.(1999)]{McHardy1999} McHardy, I.~M., Lawson, A., Newsam, A., et al., 1999, \mnras, 310, 571 

\bibitem[Mead et al.(1990)]{Mead1990} Mead, A.~R.~G., Ballard, K.~R., 
Brand, P.~W.~J.~L., et al., 1990, \aaps, 83, 183 

\bibitem[Papadakis, Villata, 
\& Raiteri(2007)]{Papadakis2007} Papadakis, I.~E., Villata, M., \& Raiteri, C.~M., 2007, \aap, 470, 857

\bibitem[Pian et al.(1999)]{Pian1999} Pian, E., Urry, C.~M., Maraschi, L., et al., 1999, \apj, 521, 112  

\bibitem[Poole et al.(2008)]{Poole2008} Poole, T.~S., Breeveld, A.~A., 
Page, M.~J., et al., 2008, \mnras, 383, 627 

\bibitem[Raiteri et al.(1998)]{Raiteri1998} Raiteri, C.~M., Villata, M., 
Lanteri, L., Cavallone, M., \& Sobrito, G., 1998, \aaps, 130, 495 

\bibitem[Raiteri et al.(2003)]{Raiteri2003} Raiteri, C.~M., Villata, M., 
Tosti, G., et al., 2003, \aap, 402, 151 

\bibitem[Raiteri et al.(2005)]{Raiteri2005} Raiteri, C.~M., Villata, M., 
Ibrahimov, M.~A., et al., 2005, \aap, 438, 39 

\bibitem[Raiteri et al.(2007)]{Raiteri2007} Raiteri, C.~M., et al. 2007, \aap, 473, 819

\bibitem[Roming et al.(2005)]{Roming2005} Roming, P.~W.~A., Kennedy, T.~E., 
Mason, K.~O., et al., 2005, Space Science Reviews, 120, 95 

\bibitem[Savolainen et al.(2002)]{Savolainen2002} Savolainen, T., Wiik, K., Valtaoja, E.,
Jorstad, S. G., \& Marscher, A. P. 2002, \aap, 394, 851 

\bibitem[Schlegel, Finkbeiner, 
\& Davis(1998)]{Schlegel1998} Schlegel, D.~J., Finkbeiner, D.~P., \& Davis, M., 1998, \apj, 500, 525 

\bibitem[Sillanp{\"a}{\"a} et al.(1993)]{Sillanpaa1993} Sillanp{\"a}{\"a}, A., Takalo, L.~O., Nilsson, K., \& Kikuchi, S., 1993, \apss, 206, 55 

\bibitem[Sokolov \& Marscher(2005)]{Sokolov2005} Sokolov, A.~S., \& Marscher, A.~P. 2005, \apj, 629, 52

\bibitem[Tornikoski et al.(1994)]{Tornikoski1994} Tornikoski, M., Valtaoja, 
E., Terasranta, H., et al., 1994, \aap, 289, 673 

\bibitem[Villata et al.(2000)]{Villata2000} Villata, M., Mattox, J.~R., 
Massaro, E., et al., 2000, \aap, 363, 108 

\bibitem[Villata et al.(2002)]{Villata2002} Villata, M., Raiteri, C.~M., 
Kurtanidze, O.~M., et al., 2002, \aap, 390, 407 

\bibitem[Villata et al.(2004a)]{Villata2004a} Villata, M., Raiteri, C.~M., 
Kurtanidze, O.~M., et al., 2004a, \aap, 421, 103 

\bibitem[Villata et al.(2004b)]{Villata2004b} Villata, M., Raiteri, C.~M., 
Aller, H.~D., et al., 2004b, \aap, 424, 497 

\bibitem[Villata et al.(2006)]{Villata2006} Villata, M., et al. 2006, \aap, 453, 817

\bibitem[Villata et al.(2007)]{Villata2007} Villata, M., Raiteri, C.~M., 
Aller, M.~F., et al., 2007, \aap, 464, L5 

\bibitem[Wehrle et al.(1998)]{Wehrle1998} Wehrle, A.~E., Pian, E., Urry, 
C.~M., et al., 1998, \apj, 497, 178 

\bibitem[Zavala \& Taylor(2004)]{ZT04} Zavala, R.~T., \& Taylor, G.~B., 2004, \apj, 612, 749

\end{thebibliography}
\end{document}